\documentstyle[twoside,fleqn,espcrc2]{article}

\newcommand{\Psibar}{\bar{\Psi}}

\newcommand{\AmS}{{\protect\the\textfont2
  A\kern-.1667em\lower.5ex\hbox{M}\kern-.125emS}}

\title{Physical Observables from Lattice QCD at Fixed
Topology\thanks{Talk presented
       by J. W. Negele. Work supported in part by the U.S. Department
       of Energy (DOE) under cooperative research agreement \#
       DE-FC02-94ER40818, DE-FG02-96ER40945 and DE-FG02-91ER40676.}}

\author{R. Brower \address {Department of Physics,
Boston University, Boston, Massachusetts 02215, USA},
 S. Chandrasekharan \address {Department of Physics,
Duke University,
Durham, North Carolina 27708, USA }
J. W. Negele \address{Center for Theoretical Physics, MIT,  77
Massachusetts
Ave.,  Cambridge, MA 02139, USA  }
and U.-J. Wiese \addressmark \thanks{ present address: University of Bern,
Sidlerstrasse 5, CH-3012 Bern, Switzerland}}

\begin{document}

\begin{abstract}
Because present Monte Carol algorithms for lattice QCD may become
trapped in a given topological charge sector when one approaches the
continuum limit, it is important to understand the effect of calculating
at fixed topology. In this work, we show that although the  restriction to
a fixed topological sector becomes irrelevant in the infinite volume
limit, it gives rise to characteristic finite size effects due to
contributions from all $\theta$-vacua. We calculate these effects and
show how to extract physical results from numerical data obtained at
fixed topology.
\end{abstract}

\maketitle

\newpage

\section{TOPOLOGICAL CHARGE SECTORS AND $\theta$-VACUA}

In principle, one should calculate observables in QCD with finite quark
masses in a fixed $\theta$-vacuum. However in practical lattice
calculations,  algorithms that change the gauge field configuration in
small steps tend to become trapped in a fixed topological charge sector
because they cannot overcome the  action barriers between sectors. One
sees concrete evidence of this problem, for example,  from the large
equilibration times for the topological charge  required in hybrid Monte
Carlo calculations, which increase as the quark mass is decreased
\cite{Bali:2001gk}. In addition, improved pure gauge actions, such as
DBW2, reduce the low eigenmodes  that are undesirable for overlap and
domain wall fermions by suppressing small instantons and thereby reduce
tunneling between sectors.  Since  we know that  the restriction to fixed
topology becomes irrelevant in the infinite volume limit, we have
calculated the the finite volume dependence of this restriction and
thereby show how to extract physical results from practical calculations
at fixed topology.

The partition function of the QCD Hamiltonian $H$ with eigenstates
$|n,\theta \rangle$ in a given sector may be written in tems of the lowest
eigenvalue $E_0(\theta) = V e_0(\theta)$ in the large volume, low
temperature limit as follows:
\begin{displaymath}
Z_\theta  = \mbox{Tr}_\theta\,  e^{- \beta H} =
\sum_n e^{- \beta E_n(\theta)}
 \rightarrow
e^{- \beta V e_0(\theta)}.
\end{displaymath}
The partition function at fixed  topological charge $Q$ is an integral
over
all $\theta$-vacuum sectors. Expanding the ground state energy
\begin{displaymath}
\label{e0chit}
e_0(\theta) = e_0(0)\;+\;\frac{1}{2}\;\chi_t \;\theta^2 \;+\;
\frac{1}{24}\;\gamma \;\theta^4 \;+\; ...,
\end{displaymath}
and performing a saddle-point expansion to ${\cal O}(1/\beta^2 V^2)$,
\begin{eqnarray}\label{ff}
Z_Q&=&\frac{1}{2 \pi} \int_{-\pi}^\pi d\theta \ Z_{\theta} e^{i \theta
Q}\\
&\rightarrow&\frac{1}{2 \pi} \int_{-\pi}^\pi d\theta \ e^{i \theta Q}
e^{- \beta V e_0(\theta)} \nonumber\\
&\simeq& e^{-\beta V e_0(0)}\sqrt{\frac{2\pi}{\beta V \chi_t}}
e^{- \frac{Q^2}{2 \beta V \chi_t}}
\nonumber \\
&\times & \!\! \!\!\!\!  { \left[1-\frac{\gamma}{24\beta V
\chi_t^2}\left( 3-\frac{6}{\chi_t}\frac{Q^2}{\beta V}+\frac{1}{\chi_t^2}
\frac{Q^4}{\beta^2V^2}\right) \right] } \nonumber
\end{eqnarray}
Using $Z_{\theta}= \int {\cal D}A {\cal D}\Psibar {\cal D}\Psi \
e^{- S[A,\Psibar,\Psi] - i \theta Q[A]} $,  the parameters in the
expansion
of $e_0(\theta)$ are
\begin{equation}\label{parms}
\chi_t = \frac{\langle Q^2 \rangle}{\beta V},\;\;\;
\gamma\;=\;
-\frac{\langle\;Q^4\;\rangle\;-\;3\langle\;Q^2\;\rangle^2}{\beta V}.
\end{equation}

The topological susceptibility,  $\chi_t$, and  fourth moment, $\gamma$,
are parameters characterizing the vacuum that can be  calculated
in lattice QCD. Once they are known, we may use eq.~(\ref{ff}) to
calculate any observable at fixed $Q$ and obtain finite volume corrections.

\section{MASS SPECTRUM IN A FIXED TOPOLOGICAL SECTOR}

As a concrete example, consider the calculation of hadron masses from
the large time behavior of two-point correlation functions. At
sufficiently low
temperature, the two-point correlation function at fixed $Q$ of operators
${\cal
O}$ with the appropriate quantum numbers to create the hadron states of
interest may be written
\begin{eqnarray*}
\label{correlation}
\langle {\cal O}(t_1) {\cal O}(t_2) \rangle_Q
&=&\frac{1}{Z_Q} \int_{-\pi}^\pi d\theta \ e^{i \theta Q
- \beta V e_0(\theta)} \nonumber \\
&\times&|\langle 0,\theta|{\cal O}|1,\theta \rangle|^2
e^{- M(\theta)(t_1-t_2)},
\end{eqnarray*}
where $M(\theta)$ is the lowest hadron mass. Performing a saddle-point
expansion, the effective mass used to measure the hadron
spectrum is
\begin{eqnarray*}
\label{MQ}
M^{eff}_Q &=& -  \frac{d}{dt_1} \log(\langle {\cal O}(t_1) {\cal O}(t_2)
\rangle_Q) \\
& = &
\frac{ \int d\theta e^{- \frac{1}{2} \langle Q^2\rangle \theta^2 +
i \theta Q} f(\theta)\;M(\theta)}
{\int d\theta e^{- \frac{1}{2}  \langle Q^2\rangle \theta^2 + i \theta Q}
f(\theta)} \; .
\end{eqnarray*}
where
$f(\theta)= |\langle 0,\theta|{\cal O}|1,\theta \rangle|^2
e^{-M(\theta)(t_1-t_2)}$  \\
contains all the time dependence. Expanding
$M(\theta) = M(0) + \frac{1}{2} M''(0) \theta^2 +
\frac{1}{4!} M''''(0)\theta^4 + \cdots $ and $ f(\theta) = f(0) +
\frac{1}{2}
f''(0) \theta^2 +
\frac{1}{4!} f''''(0)\theta^4 + \cdots $
we obtain the desired expansion,
\begin{eqnarray}
\label{MpiQ}
M^{eff}_Q &=& M(0)  + \frac{1}{2} M''(0) \; \overline{\theta^2} 
                              + \frac{1}{4!} M''''(0)\;\overline{\theta^4} \nonumber \\
&+& \frac{6f''(0)M''(0)}{4!f(0)} \left[\; \overline{\theta^4} \; - \;
(\;\overline{\theta^2}\;)^2 \right] \; ,
\end{eqnarray}
where
\begin{displaymath}
\overline{\theta^n} =
\frac{ \int d\theta \exp(- \frac{1}{2} \langle Q^2\rangle \theta^2 +
i \theta Q) \theta^n}
{\int d\theta \exp(- \frac{1}{2}  \langle Q^2\rangle \theta^2 +
i \theta Q) }.
\end{displaymath}
The moments required in eq.~(\ref{MpiQ}) are
\begin{eqnarray*}
\overline{\theta^2} &=& \frac{1}{\langle Q^2 \rangle} \left\{1  -
\frac{Q^2}{\langle Q^2 \rangle}\right\}\; , \\ \nonumber
\overline{\theta^4} &=& \frac{1}{\langle Q^2 \rangle^2}\left\{ 3 - \frac{6
\;Q^2}{\langle Q^2 \rangle} \;  +\frac{Q^4}{\langle Q^2 \rangle^2} \right\}\; ,
\end{eqnarray*}
where $\langle Q^2\rangle \;=\; \beta V \chi_t$ by eq.~(\ref{parms}).
Note that to  leading order, all dependence
of the mass shift on $f(\theta)$  exactly cancels and
therefore the result is independent of $t_1-t_2$.

Our  leading order result for the $Q$-dependence of the mass is thus
\begin{equation}\label{result}
M^{eff}_Q = M(0) +\frac{1}{2}  \frac{M''(0)}{\beta V \chi_t}
\left[ 1 - \frac{Q^2}{\beta V \chi_t} \right].
\end{equation}
Therefore, if a  calculation  in a
volume $\beta V$ is trapped in a $Q$ sector, the error is of order
$1/(\beta V)$.
Furthermore, one can measure $M_Q$ in several sectors and at
several space-time volumes and thereby extract
$M(0)$ , $M''(0)$, and $\chi_t$ using eq.~(\ref{result}).
Also note that when averaged over $Q$ with the distribution
$e^{-\frac{Q^2}{2 \langle Q^2 \rangle}}$,
$\langle \overline{\theta^2} \rangle =
0$.

\section{$\theta$ DEPENDENCE IN THE CHIRAL AND LARGE $N_c$ LIMITS}

It is possible to make some general arguments concerning the
$\theta$-dependence in QCD. The transformation  $\psi^{\prime} =
\exp(i\frac{\theta}{2 N_f}\gamma_5) \psi$ shifts the
$\theta$-dependence to the mass term,
$$m \overline{\Psi}\Psi \rightarrow  \overline{\Psi}\left\{  m\cos(\theta/N_f) +
im\sin(\theta/N_f)\gamma_5\right\}\Psi$$  and the contribution
to the fermionic measure  cancels the  $ F \tilde F$-term.
By parity, the $\sin(\theta/N_f)$ term only contributes in even orders, so
at
linear order in the quark mass, the  only effect of
$\theta$ enters through the change of the mass term to
$m\cos(\theta/N_f)$. Hence, for example, to leading order in the chiral
limit, the pion, nucleon and $\eta^{\prime}$ masses must have the forms:
\begin{eqnarray*}
M^2_\pi(\theta) & = & M^2_\pi(0) + M^2_\pi(0)[\cos(\theta/N_f)-1] \\
M_N(\theta) & = & M_N(0) +
c_1 M^2_\pi(0) [\cos(\theta/N_f)-1] \nonumber \\
&& \;\;\;\;\;\;+\; c_2 M^3_\pi(0) [\cos^{3/2}(\theta/N_f)-1] \\
M^2_{\eta^\prime}(\theta) & =  & M^2_{\eta^\prime}(0) +
b_1 M^2_\pi(0)[\cos(\theta/N_f)-1]
\end{eqnarray*}
from which the $Q$-dependence follows directly from eq.~(\ref{result}).
In the large $N_c$ limit, it can be shown that $b_1 = 1$ so that the $\pi$
and  $\eta^{\prime}$ have the same $\theta$-dependence and thus the
same $Q$-dependence.

 Using chiral perturbation theory, one can explore the chiral and large
$N_c$ limits more systematically. Using  the effective Lagrangian
proposed by Witten \cite{Witten:1980sp} to treat the $U_A(1)$ anomaly,
we obtain the following results for the $\pi$ and
$\eta^{\prime}$ masses:
\begin{eqnarray*}
M^2_\pi &=&\frac{2 m\langle\Psibar\Psi\rangle}{N_f F_\pi^2}
\cos\left(\frac{\eta^\prime_0}{F_\pi}\sqrt{\frac{2}{N_f}} -
\frac{\theta}{N_f}\right) \\
M^2_{\eta^\prime} &=&\frac{m_0^2}{N_c} +
\frac{2 m\langle\Psibar\Psi\rangle}{N_f F_\pi^2}
\cos\left(\frac{\eta^\prime_0}{F_\pi}\sqrt{\frac{2}{N_f}} -
\frac{\theta}{N_f}\right)
\end{eqnarray*}
where \\
$e_0(\theta)=-m\langle\Psibar \Psi\rangle\;
\cos\left(\frac{\eta^\prime}{F_\pi}\sqrt{\frac{2}{N_f}} -
\frac{\theta}{N_f}\right) +
\frac{m_0^2}{2 N_c}{\eta^\prime}^2.$
Solving explicitly for $ \eta^\prime_0$, we also note that in  the  limit
$ m
\rightarrow 0$, we recover $\cos\left( \frac{\theta}{N_f}\right)$  and  in
the limit $N_c \rightarrow \infty $ we obtain
$ \cos \left(\frac{m_0^2 F_{\pi}^2  \theta}{2 m
\langle \bar\psi \psi \rangle N_c}\right) \rightarrow 1$ as $N_c
\rightarrow \infty$ as expected.

\section{Q -- DEPENDENCE IN AN \\INSTANTON GAS}

Recently, the
dependence of the $\pi$ and $\eta^{\prime}$ masses on $Q$ has been
studied in lattice calculations \cite{Bali:2001gk}, which for practical
reasons, are sufficiently far from the chiral limit that the previous
results from chiral perturbation theory are inapplicable.
Each mass was evaluated in
two sectors, one with $|Q|< 1.5 $ and the other with $|Q| > 1.5$. Whereas
the pion
mass in both sectors agreed within statistics of a few percent, the
$\eta^{\prime}$ mass was of the order of 15 \% heavier in the  $|Q| > 1.5$
sector
than in the  $|Q| < 1.5$ sector.
 Motivated by the success of the Veneziano-Witten formula
\cite{Veneziano:1979ec} \cite{Witten:1979vv}  relating  the
$\eta^{\prime}$ mass to fluctuations in the topological charge, we obtain
a
qualitative understanding of the Q dependence from the fluctuations in the
topological charge arising in an instanton gas.

The vertex
generating the shift in the  $\eta^{\prime}$ mass  is proportional to the
number
of instantons,  $N$, plus the number of antiinstantons, $\bar N$. Assuming
independent Poisson distributions  with
$\langle N\rangle = \langle \bar N\rangle =
\lambda$, the probability of having $N$ instantons and $\bar N$
antiinstantons
with the constraint $N - \bar N = Q$ is:
\begin{displaymath}
P_Q(N,\bar N) = \int \frac{d \theta}{2 \pi} \frac{\lambda^{(N+\bar N)}}{N!
\bar N!} e^{-2 \lambda -i \theta(Q-N+\bar N)}
\end{displaymath}
Summing over $N$ and $\bar N$ and distinguishing $\lambda$ and
$\lambda^{\prime}$ for subsequent convenience, we write the generating
function
\begin{eqnarray*}
Z_Q(\lambda, \lambda^{\prime}) \!\!&= &\!\!\sum_{N,\bar N} \int \frac{d
\theta}{2
\pi}
\frac{{\lambda^{\prime}}^{(N+\bar N)}}{N! \bar N!} e^{-2 \lambda -i
\theta(Q-N+\bar N)}\nonumber \\
& = & \!e^{-2 \lambda} \int \frac{d \theta}{2 \pi} e^{-i \theta Q + 2
\lambda^{\prime} \cos (\theta)} \nonumber \\
& \simeq & \!\frac{1}{2 \sqrt{\pi \lambda^{\prime}}}
 \, e^{2 (\lambda^{\prime} -\lambda)}
 e^{- \frac {Q^2}{4 \lambda^{\prime}}}
\end{eqnarray*}
Differentiation of $\ln Z_Q(\lambda, \lambda^{\prime})$ with respect to
$\lambda^{\prime}$, setting $\lambda =\lambda^{\prime}$, and noting $2
\lambda = \langle Q^2 \rangle = \chi_t \beta V$ yields the desired result
for the
density of instantons plus antiinstantons at fixed Q:

\begin{equation}
\label{nplusnbar}
\frac{ \langle N+\bar N \rangle_Q}{\beta V} = \chi_t + \frac{1}{2 \beta V}
\left[
-1 + \frac{Q^2}{\chi_t \beta V} \right]
\end{equation}

 Eq. \ref{nplusnbar} is a simple and physically appealing result.  Since
$M^2_{\eta^{\prime}}
= M^2_{\pi} + \mu$ and $\mu \propto \frac{ \langle N+\bar N
\rangle_Q}{\beta V}$, which equals $\chi_t$
 when averaged over the distribution
$ P(Q) =  \frac{1}{2 \sqrt{\pi \lambda}}
 \,  e^{- \frac {Q^2}{4 \lambda}}$, the $\eta^{\prime}$ mass is consistent
with
the Veneziano-Witten formula. The finite volume corrections provide the
desired Q dependence.

To compare with the lattice results of ref
\cite{Bali:2001gk}, we note that $\beta V = 6.81 fm^4$,  $\chi_t = 0.70 fm
^{-4}$,
and $\langle Q^2 \rangle = \beta V \chi_t = 4.75$.  In the chiral limit,
$M_{\pi}^{eff} (Q) = M_{\pi}(0) - \frac{M_{\pi}}{4 N_f^2} \left[
\frac{1}{\langle
Q^2\rangle} \left( 1 - \frac{Q^2}{\langle Q^2\rangle} \right) \right] $ so
the
shift at $Q^2 = 0$ is $-M_{\pi}/(4 N_f^2 \langle Q^2\rangle ) = 0.013
M_{\pi}$.
Hence, the effect of calculation at fixed topology is of the order of 1\%,
consistent with the lattice $\pi$ results. The analogous chiral estimate for
the  $\eta^{\prime}$ is of order $(M_{\pi} /
M_{\eta^{\prime}})^2 \times 1 \%$, in strong disagreement with lattice
results.
However, using the instanton gas result, $\int dq P(Q) \frac{ \langle
N+\bar N
\rangle_Q}{\beta V}$ is 0.65 for $Q < 1.5$ and 0.75 for $Q > 1.5$,
yielding $\delta
M_{\eta^{\prime}}/ M_{\eta^{\prime}} = 8\% $, which is of the order of
magnitude of the observed $Q$ dependence.

\end{document}